\documentstyle[12pt,aasms4]{article}
\def\degr{\hbox{$^\circ$}}
\lefthead{Kalinkov et al.}
\righthead{Correlation Function of Superclusters of Galaxies}
\begin{document}
%
%\twocolumn
%

\title{Correlation function of superclusters of galaxies} \author{M.
  Kalinkov, I. Valtchanov and I. Kuneva}

\affil{Institute of Astronomy, Bulgarian Academy of Sciences, 72
  Tsarigradsko Chaussee blvd, 1784 Sofia, Bulgaria; e-mail:
  markal@astro.acad.bg, ivan@astro.acad.bg}

\begin{abstract}
  We present a study of the two-point correlation function $\xi(r)$ of
  superclusters of galaxies. The largest catalogs are used. The
  results show negligible correlation $|\xi| < 0.1 \div 0.2$ for
  separations up to $(500\,\div\,600)\ h^{-1}$ Mpc. Small correlations
  are obtained using various estimates and samples.  Seemingly there
  are no structures of superclusters of galaxies.
\end{abstract}


\section{Introduction}
Correlation functions are very useful instrument to study the
structuring in the Universe (Peebles 1980 and references therein).
Despite some discussion on the dependence of the correlation radius on
depth, luminosity or richness, the two-point correlation function for
galaxies and clusters is well established -- a power low $\xi(r) =
(r/r_0)^{-\gamma}$, where $\gamma\approx 1.7\, \div\, 1.8$ and $r_0$
is the correlation radius (see Bahcall 1988 for a review; Fisher et
al. 1994, Loveday et al. 1995, Hermit et al. 1996, Postman et al.
1986, 1992, West \& van den Bergh 1991, Efstathiou et al. 1992). As
concerning the superclusters of galaxies Kalinkov \& Kuneva (1985)
have shown that $\xi(r h^{-1}) \approx 0$ for $r \la 180 h^{-1}$ Mpc
($h$ is the Hubble constant in units of 100 km s$^{-1}$ Mpc$^{-1}$; we
use luminosity distance for $q_0$ = 1/2).  This result was confirmed
later by Kalinkov \& Kuneva (1986).  Bahcall \& Burgett (1986)
revealed correlations for superclusters from the catalog of Bahcall \&
Soneira (1984) on a scale of $(100\,\div\,150)\ h^{-1}$ Mpc,
significant at $3\sigma$ level. So a universal dimensionless
correlation function was proposed -- for galaxies, clusters and
superclusters (cf.  Szalay \& Schramm 1985). Lebedev \& Lebedeva
(1988) found a power low correlation function with $\gamma$ just 1.8
for the catalog of Batuski \& Burns (1985). But the result is doubtful
because selections in the real catalog were not taken into account.
Besides, this supercluster catalog has large uncertainties since very
rough redshift estimates were used.

We have obtained new estimates for the space two-point correlation
functions on the base of the five most complete recent catalogs of
superclusters of galaxies.

In section 2 we describe the catalogs, in section 3 our procedures for
calculation of the correlation function are presented.  The results
are given in section 4 and are discussed in section 5.

\section{Catalogs of superclusters of galaxies}

All current catalogs of superclusters (with an only exception, see $\S
2.2.3$) are compiled from the A-ACO catalogs of clusters of galaxies
(Abell 1958, Abell et al. 1989). In fact the ACO catalog contains all
the data from the original Abell catalog.

\subsection{Previous lists}

First lists of superclusters in the redshift space are published by
Thuan (1980) and Bahcall \& Soneira (1984). Only A-clusters with
richness ${\cal R} \geq 1$ are used.  Thuan have examined 77 clusters
with measured redshift $z_{m} < 0.08$ and found 17 superclusters with
multiplicity $2 \leq \nu \leq 11$ using a percolation procedure with
percolation radius $r_{p} = 36h^{-1}$Mpc.  Bahcall and Soneira used a
nonpercolation as well as a percolation procedure and found 16
superclusters (from 104 A-clusters with $z_{m} \la 0.1)$ at density
enhancement $f = 20$ with $2 \leq \nu \leq 15$, while at $f = 400$
there are only 7 with $\nu = 2$ or 3. The corresponding $r_{p} =
14h^{-1}$Mpc for $f = 20$. Note the quantity $f$ has different meaning
in various publications.

The finding list of candidate superclusters of Batuski \& Burns (1985)
includes 102 aglomerations with $r_{p} = 30h^{-1}$Mpc among 652
A-clusters within $z = 0.13$. But the larger part of the clusters have
estimated redshift $z_{e}$ with a standard deviation of 30 \%. Some
of the superclusters are extremely huge and contain tens of clusters.
Tully (1987) also found huge superclusters. The existence of such vast
complexes was put in question (e.g. Postman et al. 1989, 1992).

West (1989) found 48 superclusters with $r_{p} = 25h^{-1}$Mpc
(corresponding to $f = 11$) among 286 A-clusters with ${\cal R} \geq
0$ within $z_{m} = 0.1$. The multiplicity is from 2 to 13.

Postman et al. (1992) studied the distribution of near A-clusters (a
complete sample of 351 clusters with $z_{m}$ up to $m_{10} = 16.5$)
and with $r_{p} = 22$, 16, and 13$h^{-1}$Mpc (equivalent to $f = 2$,
5, and 10) they have found 23 superclusters at $f = 2$ with $\nu \geq
3$, while there are 11 superclusters at $f = 10$ with $\nu = 3$ or 4.

Another percolation analyses of near A- and ACO-clusters ($z \la 0.08,
{\cal R} \geq 0$) with only 14 estimated redshifts (acccording to
Scaramella et al. 1991, with st.dev. of 34~\% for A- and 25 \% for
ACO-catalog) is performed by Cappi \& Maurogordato (1992). At density
enhancement $f = 1.9$ for northern clusters $r_{p} = 21h^{-1}$Mpc
while for southern ones $r_{p} = 22h^{-1}$Mpc. There are 24
superclusters having $\nu \geq 3$ at $f = 1.9$.

\subsection{Current catalogs}

Three catalogs of near superclusters, one catalog of near and distant
superclusters, and the newest catalog of APM superclusters are used in
our analysis.

\subsubsection{Three catalogs}

{\bf ZZSV} -- Zucca et al. (1993) found superclusters among A-ACO
clusters having richness ${\cal R} \geq 0$ and distance $R <
300h^{-1}$Mpc.  ZZSV used redshift-magnitude relations to estimate
redshift of A- and ACO-clusters with st.dev. of 23 \% and 24 \%
respectively. The percolation radii are different for A- and
ACO-clusters -- for $f \geq 2$ (their denotation) $r_{p} = 20.1$ and
16.8$h^{-1}$Mpc, while at $f \geq 200$ they have $r_{p} = 4.3$ and
3.6$h^{-1}$Mpc. ZZSV found $N = 69$ superclusters with $\nu \geq 3$
for $f \geq 2$.

{\bf EETDA} -- Einasto et al. (1994) used $r_{p} = 24h^{-1}$Mpc for
783 A-ACO clusters (${\cal R} \geq 0, R \leq 300h^{-1}$Mpc) and found
$N = 130$ superclusters with $2 \leq \nu \leq 32$. Estimated redshifts
are according to the magnitude-redshift relations in the ACO catalog
(with error $ > 40 \%$) or according to Postman et al. (1985), with
st.dev. of 38 \% .

{\bf ETJEA} -- the last catalog of Einasto et al. (1997) compiled
again with $r_{p} = 24h^{-1}$Mpc) among 1304 A- and ACO- clusters
$({\cal R} \geq 0,\,z \leq 0.12)$. Here $z_{e}$ is estimated according
to Peacock \& West (1992) with st.dev of 27 \% for A-clusters and 18
\% for ACO-clusters.  The multiplicity is $2 \leq \nu \leq 34$ for all
of the 220 superclusters.

\subsubsection{An extensive catalog}

The largest catalog of superclusters among all A- and ACO-clusters is
compiled by Kalinkov \& Kuneva (1995) -- KK, with a nonpercolation
procedure, for a local density enhancements $f = 10,20,40,100,200$ and
400. However a resemblence between our procedure and the percolation
analysis allows to define an acceptable $r_{p}$ depending on distance,
for both procedures.  Thus $r_{p} = (7.0 \div 10.4)h^{-1}$Mpc for $f =
10$ and $r_{p} = (2.1 \div 3)h^{-1}$Mpc for $f = 400$.

There are 893 different superclusters. The catalog contains
superclusters with members with measured ($z_{m}$) and/or estimated
($z_{e}$) redshift. The estimated redshifts are computed according to
the multiple regressions found by Kalinkov et al. (1994). We have to
note that the st.devs of the regressions for A-clusters are: 4.4 \%
for ${\cal R} = 0 + 1$, 3.3~\% for ${\cal R} = 2$ and 1.9 \% for
${\cal R} = 3$. Small st.devs for A-clusters is expectable since many
good regressors are used. The redshift estimates are not so good for
ACO-clusters -- $s = 33 \% $ for ${\cal R} = 0, s = 20 \% $ for ${\cal
  R} = 2$ and $s = 18 \%$ for ${\cal R} = 3 + 4.$ Our standard
deviations for ACO-clusters are worse than those of Peacock \& West
(1992). The cause may be that Peacock \& West rejected many measured
redshifts. The rejection of values which strongly depart from a
regression curve inevitably leads to decreasing of the standard
deviation of the regression. The absence of good regressors influences
our st.dev.

Some general characterisrics of the superclusters are given in KK,
kinematical and dynamical features may be found in Kalinkov et al.
(1996) and the distribution of the superclusters in Kalinkov et al.
(1998).

We give in Table 1 the most representative samples of our catalog --
for $|b| > 30\arcdeg,\, \nu \geq 3,\,100 \leq Rh^{-1}{\rm Mpc} \leq
627$, i.e. up to $z = 0.20$, and for $f = 10$ and 20. All member
clusters have measured redshift.

%\placetable{tbl-1}

The columns contain: the number in the KK catalog, the local density
enhancement $f$, cluster members, identifications with superclusters
from previous lists or catalogs and notes. The case $f = 10;20$ means
that the membership at $f = 20$ is different at the lower enhancement,
and supercluster members for $f = 20$ are bold. The semicolumn
discriminate the identifications for $f = 10$ and 20. All lists and
catalogs are shortened: T - Thuan, B - Bachall \& Soneira, P --
Postman et al., C - Cappi \& Maurogordato, Z - Zucca et al., E - EETDA
and ET - ETJEA. When it is necessary, the corresponding density
enhancement is given after a slash . Cappi \& Maurogordato work at $f
= 1.9$, but for brevity we denote it with 2.

Thus all superclusters in the column ``Identifications'' are just
equal to the superclusters from the first column. The sign ``$+$'' or
``$-$'' shows which ACO cluster have to be added or removed from the
configuration to reach an equallity. The cases when the equallity
could be reached with addition or removal of more than two
ACO-clusters are given in the notes. The corresponding multiplicity
is given in brackets.

Table 1 shows that the larger part of our superclusters, independently
from the searching procedures, have entries in the other lists or
catalogs. Only distant superclusters have no correspondence since the
other catalogs are delimited at about 300$h^{-1}$Mpc.  Thirteen
superclusters in Table 1 are entirely new ones.  The older lists do
not resemble the superclusters from Table 1 well, because they have
studied only clusters with richness ${\cal R} \geq 1$.

Table 2 contains more data for the superclusters from Table 1. Columns
1-9 contain supercluster designation, density enhancement $f$, multiplicity
$\nu$, coordinates \mbox{(HHMM.d $\pm$ DDMM)}, galactic latitude $b$,
distance $R$, the extension along R.A., Dec, and R. Next columns 10-13
give four space radii -- inertial $r_{i}$, mean harmonic weighted
$r_{h}$, mean separation $r_{s}$, virial $r_{v}$ and the last column
is the mass, defined as simple sum of cluster masses, computed
according to the relation of Bahcall \& Cen (1993).

In fact we have used ten radii -- from projected distribution (2D) and
for space (3D) case, as well as weighted and unweighted ones. All the
radii are in accordance implying that there is no significant
anysothropy in the cluster distribution inside the superclusters.
effects. The radii we have used are the same ones used in galaxy
groups studies (the formulae are given by Huchra \& Geller 1982,
Nolthenius \& White 1987, Ramella et al. 1989, Maia et al. 1989,
Gourgoulhon et al. 1992 and Jackson 1975).

First of all, the superclusters from Tables 1--2 are not elongated in
the line of sight direction. For $f = 10$ we have $<\Delta {\rm
  R.A.}>:<{\rm Dec}>:<R>=(14.0\pm8.8): (15.6\pm7.8):(17.5\pm9.1)$ and
$(9.8\pm6.5):(11.6\pm6.3): (13.2\pm7.3)$ for $f = 20$ (all extensions
are in $h^{-1}$Mpc). This fact leads to the conclussion that the
peculiar velocities of the clusters are not larger than a few hundred
km s$^{-1}$.

The inertial radius $r_{i}$ is the smallest -- $<r_{i}> =
11.7\pm3.7h^{-1}$Mpc, while the virial one is the largest -- $<r_{v}>
= 22.5\pm10.2h^{-1}$Mpc for $f = 10$. 

The radii regarded as physical size of the superclusters put
constraint on the smallest separation for correlation function
estimates $\Delta r \ga 40h^{-1}$ Mpc.

The mean supercluster mass for $f = 10$ is $(1.36\pm0.50) \cdot
10^{15} {\cal M}_{\sun}$.  The mass for $f = 20$ is somewhat smaller
-- $(1.30\pm0.42) \cdot 10^{15} {\cal M}_{\sun}$.

\subsubsection{The APM catalog}

The last catalog of clusters of galaxies (Dalton et al. 1997) is
constructed on the APM Galaxy Survey, which contains over $2 \cdot
10^{6}$ galaxies to a magnitude limit of $b_{j} = 20.5$ (e.g Maddox et
al. 1990a,b; 1996). The survey covers an area of 4300 $\sq\arcdeg$ in
the southern sky where the galactic absorption is negligible. The APM
catalog of clusters of galaxies is compiled with an automated
algorithm of Abell-like selection criteria. This catalog is entirely
objective. It does not comprise any subjective or systematic errors
inherent to A- and ACO-catalogs.

We have added more measured redshifts to the APM catalog of clusters
of galaxies, taken predominantly from the literature. Thus the number
of clusters with measured redshifts in our version of the catalog is
393. A new estimate of the redshift for the rest of 564 clusters is
obtained with the regression
\begin{flushleft}
\begin{displaymath}
\arraycolsep = 1pt
\begin{array}{rcrlcrl}
\log z_{e} = -4.36  &+& 0.176  & m_{X} &+& 0.068  & \log(\pi r_{C}^{2}/N),\\
             \pm 44 & & \pm 23 &       & & \pm 47
\end{array}
\end{displaymath}
\end{flushleft}
where $r_{C}$ is the projected radius used in the original catalog in
the final iteration and $\cal N$ is the cluster richness (the notation
in Dalton et al. 1997 is $\cal R$). The standard deviation for the
above regression is 28 \% .

We searched the APM catalog of clusters for superclusters of galaxies
with a percolation procedure and $r_{p} = 15h^{-1}$Mpc, which
corresponds roughly to density enhancement $f = 10$. There are 83
superclusters with multiplicity $\nu \geq 3$. The sixteen
superclusters whose members have measured redshift are given in Table
3. In brackets is given the rich cluster (A) or the supplementary (S)
cluster which is identified with APM cluster. It is curious that
suplementary clusters, which are regarded by Abell et al. (1989) as
poor or very distant, are found in an objective searching procedure.
It may be caused by the background correction accepted in ACO, based
on a ``universal'' luminosity function for galaxies. It is not by
chance that in some cases in ACO ${\cal R} < 0$ (S17, S34 -- their
Table 5; A2604, A2655 -- their Table 6). The original procedure of
Abell (1958) require a local correction.  Actually the APM catalog of
clusters is based on a local background correction.  So some of the
supplement ACO clusters could be rich and near ones.  The nearest
APM supercluster is at $R = 149h^{-1}$Mpc and the distant one -- at
$361h^{-1}$Mpc.

\section{Estimates and uncertainties of the correlation function}

We use three estimates of the correlation function:

DP (Davis \& Peebles 1983):
\begin{displaymath}
\xi(rh^{-1}) + 1  =  \frac{2~DD}{DR}\frac{n}{n-1},
\end{displaymath}

H (Hamilton 1993):
\begin{displaymath}
\xi(rh^{-1}) + 1  = \frac{4~DD~RR}{DR^{2}}\frac{n^{2}}{(n-1)^{2}}\ \ {\rm
  and}
\end{displaymath}

LS (Landy \& Szalay 1993):
\begin{displaymath}
\xi(rh^{-1}) + 1  = \left(DD - \frac{DR}{2}\frac{n-1}{n} + RR\right)/RR,
\end{displaymath}
where $DD$ is the number of pairs in the separation bin $(r-\Delta
r/2,\ r+\Delta r/2)$ in a supercluster catalog sample, $RR$ is the
number of pairs in a random catalog with same number of objects $n$,
occupying the same volume and with the same selections as the real
one, and $DR$ is the crosscorelation pair counts between both
catalogs.

We have used three uncertainties in terms of significant interval
corresponding to $1 \sigma$ (68 \%):

P (Poisson):
\begin{displaymath}
\Delta \xi_{P} = \frac{1 + \xi}{\sqrt{DD}},
\end{displaymath}

LFB (Ling, Frenk \& Barrow 1986):
\begin{displaymath}
\Delta \xi_{L} = \left[\Sigma^{N}_{i=1}(\xi_i -
  <\xi>)^2/(N-1)\right]^{1/2},
\end{displaymath}
with $<\xi> = \Sigma^{N}_{i=1}\xi_i/N$, where $\xi_i$ is a bootstrap
estimate and $N$ is the number of mock catalog generations; that is
the uncertainty obtained with the bootstrap resampling technique, and

ET (Efron \& Tibshirani 1986): $\Delta \xi_{ET}$,\\
defined from the bias-corrected 68 \% confidence interval from
\begin{displaymath}
\xi_{BC}(t) = G^{-1}\left[\Phi\left(\Phi^{-1}(t)
  + 2 \Phi^{-1}\left(G(<\xi>)\right)\right)\right],
\end{displaymath}
where $G$ is the cumulative distribution function (CDF) of the
$\xi_{i}$'s, and $\Phi$ is the CDF for the normal distribution (thus
$\Phi^{-1}(t) = -1,\ +1$). So $\Delta \xi_{ET}$ is the bootstrap
uncertainty.

There are plenty of opinions for the uncertainties of the correlation
function. Peebles (1980) argue that its variance is $1/DD$, namely
Poissonian, if the correlation is negligible. Landy \& Szalay (1993)
demonstrated that $1/DD$ variance is not attained for many estimators.
Their estimator (LS) is free of biases and has $1/DD$ variance {\em at
  all scales} for uncorrelated data. (A treatment for the variance of
the correlation function in the presence of correlations is given by
Bernstein 1994).  According to Ling et al.  (1986) $\Delta \xi_{P}$
underestimates the uncertainties, while Mo et al.  (1992) reckon that
the bootstrap resampling technique overestimates the uncertainties in
the correlation function.  A detailed comparison of various estimators
for the correlation function could be found in Sicotte (1995). An
extremely successful application of $\Delta \xi_{ET}$ is made by
Shepherd et al. (1996).

Usually the uncertainty $\Delta\xi_{L}$ is called ``bootstrap'' (e.~g.
Mo et al. 1992). We prefer using this term to $\Delta\xi_{ET}$.

Some information about the random catalogs. Space coordinates of any
supercluster are defined by the galactic coordinates $l,\ b$ and
distance $R$. Applying the bootstrap method we choose randomly
from the real superclusters a latitude, a longitude and a distance
independently, thus forming one random object. In all estimations we
generate 1000 random catalogs with number of objects equal to the
number of objects in the supercluster catalog sample.

We have probed another way to construct the random catalog. Knowing
the distance and latitude selection functions, and supposing a random
distribution along longitude, it is easy to generate a catalog which
follows the selections in the real one.

It is worth to note that when the bootstrap resampling technique for
the uncertainties is used then the random catalog is constructed by
picking up randomly objects from the supercluster catalog sample. A
detailed description is given by Mo et al. (1992).

\section{Results}

We present results for nine samples (Table 4) from four catalogs in
the polar caps $|b| \geq 30\degr$. While ``N'' denotes the northern
cap, ``N+S'' means joint examination of the superclusters in both
caps. Only superclusters having multiplicity $\nu \geq 3$ are
examined.  The number of objects is $n$. In column ``Distance'' $z$
means that all member clusters of the superclusters have measured
redshift.  Sample 4 contains superclusters with at least one member
cluster having measured redshift denoted as ``$z+(e)$''. For samples
5, 6 and 7 ``$z+e$'' means that the superclusters have measured as
well as estimated redshifts.

Samples 1-4 contain superclusters in distance interval $100 \leq
R~h^{-1}~{\rm Mpc} \leq 627$, where 627 $h^{-1}$ Mpc corresponds to $z
= 0.20$. The superclusters from sample 5 are in the interval $43 \leq
R~h^{-1}~{\rm Mpc} \la 300$. A direct comparison between distances for
common superclusters in KK and EETDA as well as ETJEA shows that in
both last catalogs a proper-motion distance is used, which is the
luminosity distance divided by $(1+z)$. Thus sample 6 contains
superclusters in the interval $55 \leq R~h^{-1}~{\rm Mpc} \la 300$,
while sample 7 -- $61 \leq R~h^{-1}~{\rm Mpc} \la 330,$ but here $R$
is the proper-motion distance.  Sample 7 contains only the real
clusters and not the supercluster candidates as defined by ETJEA.
Samples 8-9 include the APM superclusters with $149 \leq Rh^{-1}{\rm
  Mpc} \leq 348.$

Figs. 1-3 contain correlation functions according to the three
estimators. The uncertainty $\Delta \xi_{L}$ is associated with open
and full circles as well as diamonds located at the volume center of
the corresponding bin. Poisson uncertainties $\Delta \xi_{P}$ for
clarity are given on the left, while the bootstrap uncertainties are
shifted to the right of their corresponding place.

All estimates in the first bin are biased as far as the superclusters
are not points, while the objects in the random catalogs are points.
(When ``sizes'', e.g. virial radii, are attributed to the random
``superclusters'' then the estimates in the first bins will be
unbiased). The bias slightly depends on the sample size as well as on
the bin widths. However we present the biased estimates.

The correlation functions for samples 1--3 are the most representative
ones. All member clusters of the superclusters have measured (not
estimated) redshift. It is evident that $|\xi| < 0.1 \div 0.2$ for $r
\la 500~h^{-1}$ Mpc. The correlation functions for both galactic caps
present the results for a sample with a total length of $\ga
1200~h^{-1}$ Mpc.

%\placefigure{fig1}
%\placefigure{fig2}
%\placefigure{fig3}
%\placefigure{fig4}

The Poissonian error, as well as $\Delta \xi_{L}$, underestimates the
real uncertainty. It seems that the uncertainty $\Delta\xi_{ET}$ is
representative.  Another note is that the uncertainty for the LS
estimator is not a Poissonian one.

Sample 4 is not so representative as far as it includes cluster
members with estimated redshift. However this sample supports the case
for $|\xi| \la 0.1$.

Samples 5--7 have large uncertainties. In the last two samples only a
rough estimates for redshift is made. Nevertheless Fig. 2 supports the
case of small correlation.

A crucial test is based on samples 8--9, which are constructed from an
entirely objective catalog of clusters of galaxies (APMC). Fig. 3
unambiguously indicates $|\xi| < 0.1 \div 0.2$.

%\placefigure{fig5}
%\placefigure{fig6}
%\placefigure{fig7}

We show the three estimators -- DP, H, LS are in very good agreement.
All samples support our claim for non significant correlation up to $r
\approx 500h^{-1}$ Mpc.

The most astounding thing is that for all superclusters in KK catalog,
disregarding the density enhancement, multiplicity and measured or
estimated redshift, we get small correlation. This refers to various
samples from the catalogs ZZSR, EETDA, ETJEA and APMS as well.

We have determined the galactic latitude selections as well as the
distance selections for superclusters in KK catalog (Kalinkov et al.
1998).  The construction of random catalogs having the same selections
as in the real catalog does not change the main result -- negligible
correlation.

\section{Discussion}

Using some representative samples of superclusters of galaxies we have
shown that $|\xi| (rh^{-1}) < 0.1 \div 0.2$ up to separation
$500h^{-1}$ Mpc.  This is valid for a volume with length $\sim
1200h^{-1}$ Mpc for $|b| \geq 30\degr$. Five catalogs of superclusters
of galaxies are used.  Three different searching algorithms are
applied to form these catalogs. Our result is independent of the
samples -- superclusters having $\nu \geq 2,\ 3$ and 4, defined at
density enhancements $f = 10,\ 20,\ 40$ and 100 (there is no
statistics for higher $f$), including superclusters with member
clusters without measured redshift. All samples lead to negligible
correlation which may be regarded as a robust result. It seems that
the superclusters of galaxies compose a Poissonian field without any
structures of superclusters.  Therefore the superclusters do not obey
the universal correlation function hypothesis (Bahcall 1988, Bahcall
\& West 1992).

One could assume that the result of Bahcall \& Burgett (1986) for
non-zero correlation is due to the small statistics.

We thank to M. Postman, J. Huchra and W. Keel for furnishing us with
new redshifts before publication. We are grateful to G. Dalton and H.
Andernach, which place at our disposal data before publication.  We
acknowledge H. Sicotte for some discussions.

This research has made use of the NASA/IPAC Extragalactic Database
(NED) which is operated by the Jet Propulsion Laboratory, Caltech,
under contract with the National Aeronautics and Space Administration.

This work was supported by the National Research Fund of the Bulgarian
Ministry of Education, Science and Technology (contract F469/1994).

\clearpage

\figcaption[]{Correlation functions for samples 1-4 (Table 4).
  Estimators: DP -- Davis \& Peebles, H -- Hamilton, LS -- Landy \&
  Szalay. The uncertainties associated with the central estimate for
  each bin is according to the bootstrap resampling technique, while
  the Poisson uncertainties are on the left and the bootstrap ones --
  on the right. For sample 2, first bin, H estimator, the uncertainty
  $\Delta \xi_{ET} = \pm0.60.$ For sample 3, first bin, LS estimator,
  the uncertainty $\Delta \xi_{L} = \pm 0.67.$ For sample 4, first
  bin, $\Delta \xi_{ET} = \pm 0.56$ (DP), $\Delta \xi_{L} = \pm 0.73$
  and $\Delta \xi_{ET} = \pm 0.76$ (H).
\label{fig1}}
\figcaption{Correlation functions for samples 5-7 (see Fig.~1).
\label{fig2}}
\figcaption{Correlation functions for samples 8 and 9 (see Fig.~1).
  The uncertainties for the last bin of LS estimator are
  $\Delta\xi_{L} = \pm 0.49$ and $\Delta\xi_{ET} = \pm
  0.43$.\label{fig3}}

\clearpage
\begin{deluxetable}{rcrrrrrrlc}
\label{tbl-1}
\tabcolsep=0.1pt
\tableheadfrac{0.1}
\tablecolumns{10}
\tablenum{1}
\footnotesize
\tablecaption{SUPERCLUSTERS WITH MEASURED REDSHIFT ($\nu \ge 3,\,
  |b| \ge 30\arcdeg,\, 100 \le R h^{-1} {\rm Mpc} \leq 627$)}
                                %\tablewidth{0pt}
\tablehead{\colhead{SC} & \colhead{$f$}
  & \multicolumn{6}{c}{ACO-clusters}
  & \colhead{Identifications} & \colhead{Notes}}
\startdata
43&   10 &   40&   79&   84&   98&     &     & E 9$-$A40$-$A79$=$ET 13$-$A40$-$A79             &            \nl
52&   20 &   74&   80&   86& 2800&     &     & P 13a/10$+$A80$=$C S1a/5/10$+$A80$=$Z 1/10    & 1,2        \nl
64&   10 &  102&  116&  134&     &     &     & W 20$=$P 15/2/5$=$C S3/2$=$Z 5/2$=$E 12$=$ET 19         &            \nl
67&   10 &   85&  117&  126&  151&     &     & T 1$+$ A117$+$A126$=$P 14/2$+$A126$=$C S2/2$+$A126$=$Z 4/2$+$A126    & 1,3,4      \nl
75& 10-20& 2841& 2854& 2889&     &     &     & Z 47/2$-$A2836e$=$E 10$-$A2764$=$ET 21        &            \nl
79& 10-20&  119&  147&  168&     &     &     & Z 3/10$=$P 16/5$=$C S4/5/10               & 5,6,7      \nl
89& 10-20&  150&  154&  158&  171&     &     & P 17/5$+$A150$=$C S5/2$+$A150               & 8,9,10,11  \nl
96&   10 &  160&  193&  195&     &     &     &                                       & 6,7        \nl
117&   10 &  216&  217&  243&     &     &     & ET 31$-$A229                            & 12         \nl
121&   10 &  226&  228&  259&     &     &     &                                       & 12         \nl
130&   10 &  225&  257&  292&  311&     &     &                                       & 11,13      \nl
131& 10-20&  266&  277&  281&     &     &     & W 42$+$A266                             & 14         \nl
159& 10-20& 2988& 3004& 3009&     &     &     &                                       & 15         \nl
203& 10-20&  419& 3094& 3095&     &     &     & P 18/2$-$A428$=$P 18/5/10$=$C S6/2$-$A428$=$C S6/5$=$Z 8/2$-$A428   & 16         \nl
&      &     &     &     &     &     &     & $=$ET 49$-$A428$-$A3151 &   \nl
206& 10;20& 3093& {\bf 3100}& {\bf 3108}& {\bf 3109}&     &     & Z 21/10$-$A3104$+$A3109              & 16,17,18   \nl
225& 10;20& {\bf 3125}& {\bf 3128}& {\bf 3158}& 3164&     &     & C S11b/5; C S11b/10                   & 16,18,19   \nl
241&   10 & 3144& 3193& 3202&     &     &     & C S12/2$-$A3225$=$C S12/5$-$A3225$=$Z 50/2$=$E 29$-$A3144& 20         \nl
258& 10-20& 3225& 3231& 3266&     &     &     &                                       & 21         \nl
270&   10 &  484&  496&  536&     &     &     & W 33$+$A536$=$E 31$+$A536                   &            \nl
315& 10-20&  690&  692&  699&     &     &     & Z 12/2$=$E 47$-$A722$=$ET 76$-$A722           &            \nl
332&   20 &  762&  765&  787&  788&     &     &                                       & 12         \nl
343& 10;20&  {\bf 786}&  809&  {\bf 818}&  {\bf 848}&     &     &                                       & 12         \nl
377&   20 &  950&  985& 1002&     &     &     &                                       & 12         \nl
378& 10;20&  {\bf 930}&  {\bf 970}&  {\bf 978}&  {\bf 979}&  {\bf 993}& 1069& W 15$+$A930$+$A970$=$P 1/2$=$Z 14/2$=$E 53;     &            \nl
&      &     &     &     &     &     &     & P 1/5$=$Z 5/10$+$A970$=$ET 88$+$A930           &            \nl
379& 10-20&  965&  980& 1000&     &     &     &                                       & 12         \nl
419&   10 & 1155& 1187& 1190&     &     &     & W 47$+$A1155$=$E 55$-$A1097$=$ET 95$-$A1203     &            \nl
420&   10 & 1149& 1171& 1238&     &     &     & P 3/2$=$C N2/2$=$E 56$-$A1066               & 22         \nl
431& 10-20& 1177& 1185& 1228& 1257& 1267&     & P 4/2/5$=$C N3/2/5$=$Z 18/2             & 23         \nl
435& 10;20& {\bf 1123}& {\bf 1150}& {\bf 1297}& {\bf 1301}& 1381&     &                                       & 12         \nl
448& 10-20& 1216& 1308& 1334&     &     &     & W 22$=$P 5/2$=$C N4/2$=$Z 19/2$=$E 59$=$ET 98   &            \nl
462&   10 & 1239& 1357& 1359&     &     &     &                                       & 12         \nl
470& 10-20& 1341& 1342& 1345& 1372&     &     &                                       & 24         \nl
473& 10-20& 1307& 1337& 1385& 1390&     &     & Z 21/2                                & 25         \nl
478&   10 & 1218& 1400& 1468&     &     &     & E 60$+$A1468$=$ET 99$+$A1468                & 26         \nl
483& 10-20& 1302& 1366& 1406& 1421& 1432&     &                                       & 12,27      \nl
486& 10;20& {\bf 1291}& {\bf 1318}& {\bf 1377}& 1383& 1436& 1452& T 9$+$A1452$+$A1507$=$B 8/20/40/100$+$A1452$+$A1507$=$P 6/2$-$A1270           &            \nl
&      & 1507&     &     &     &     &     & $=$P 6/5$+$A1507$=$C N5/2$-$A1270$=$C N5/5$+$A1507$=$Z 20/2;         & 26,28    \nl
&      &     &     &     &     &     &     & B 8/200/400$-$A1383$+$A1377$=$P 6/10 $-$A1383$=$C N5/10$+$A1383          &            \nl
&      &     &     &     &     &     &     & $=$Z 7/10$-$A1383$=$E 66$-$A1270$=$ET 109$-$A1270               &            \nl
493& 10-20& 1365& 1423& 1480&     &     &     & E 67                                  & 29         \nl
524&   10 & 1474& 1526& 1552& 1569&     &     & W 12$+$A1526$=$Z 23/2$-$A1541$-$A1589         & 30         \nl
537&   10 & 1559& 1561& 1597& 1674&     &     &                                       & 12,31      \nl
543&   20 & 1566& 1621& 1646&     &     &     &                                       & 12,27      \nl
558& 10-20& 1661& 1667& 1679&     &     &     &                                       & 12         \nl
560&   10 & 1672& 1675& 1677&     &     &     &                                       & 12         \nl
587& 10-20& 1773& 1780& 1809&     &     &     & T 11$+$A1780$=$B 11/20$\div$400$+$A1780$=$W 32$+$A1780$=$Z 26/2 &            \nl
&      &     &     &     &     &     &     & $=$E 82$=$ET 136$-$A1784            &            \nl
595&   20 & 1825& 1827& 1828&     &     &     &                                       & 32,33      \nl
598&   10 & 1775& 1800& 1831& 1873& 1898&     & P 7/2$\div$10$+$A1898$=$C N7/2$\div$10$+$A1898$=$Z 27/2$+$A1831$+$A1898            &  33,34     \nl
623&   10 & 1920& 1936& 1937& 1940&     &     &                                       & 12         \nl
624&   10 & 1899& 1913& 1991&     &     &     & E 84$-$A1991$=$ET 143$-$A1991               & 35         \nl
635& 10-20& 1925& 1962& 1999& 2000&     &     & E 86$=$ET 147                           &            \nl
640& 10-20& 1972& 1976& 1980& 1986& 1988& 2006&                                       & 12,36      \nl
641& 10-20& 1984& 1990& 2005&     &     &     &                                       & 12         \nl
645& 10-20& 2001& 2008& 2017&     &     &     &                                       & 12,37      \nl
648&   20 & 2028& 2029& 2033&     &     &     & T 13$+$A2033$=$W 36$+$A2033$=$Z 30/2/10$=$E 89$-$A2040$=$ET 154$-$A2066             &    38      \nl
661& 10;20& {\bf 2061}& {\bf 2065}& {\bf 2067}& 2079& {\bf 2089}& 2092& T 15$-$A2124$=$B 12B/20$\div$100$-$A2124$=$P 10/2$-$A2124              &  39,40     \nl
&      &     &     &     &     &     &     & $=$C N10/2$-$A2124$=$ET 158$-$A2019$-$A2124;    &            \nl
&      &     &     &     &     &     &     & B 12B/200$=$P 10/5$=$C N10/5              &            \nl
665& 10-20& 2093& 2096& 2100&     &     &     &                                       & 12         \nl
677&   20 & 2107& 2147& 2148& 2151& 2152&     & B 15/20$-$A2063$+$A2148$=$B 15/40$+$A2148$=$P 9/5$=$C N9/5             & 41,42      \nl
&      &     &     &     &     &     &     & $=$Z 13/10$+$A2107$+$A2148  &           \nl
679&   10 & 2142& 2175& 2178&     &     &     & B 12C/20/40$+$A2178                     &            \nl
681& 10-20& 2168& 2169& 2184&     &     &     & P 11/2$=$C N11/2$=$Z 32/2$=$E 93$-$A2149  &            \nl
683& 10-20& 2158& 2172& 2179& 2183& 2196& 2211&                                       & 12         \nl
699&   10 & 2245& 2249& 2253&     &     &     & E 95$+$A2253$=$E 167$+$A2253                &            \nl
702& 10-20& 2248& 2256& 2271& 2296&     &     & W 18$+$A2296$=$P 12/2$-$A2309$=$P 12/5/10$+$A2248$=$Z 33/2$-$A2295   &            \nl
&      &     &     &     &     &     &     & $-$A2309$=$Z 14/10/20$+$A2248$=$E 96$-$A2309$=$ET 168$-$A2309     &            \nl
755& 10;20& {\bf 2361}& {\bf 2362}& 2372& {\bf 2382}& 2401&     & Z 35/2;                                & 43         \nl
763& 10-20& 3806& 3822& 3825& 3826&     &     & C S13/5/10$=$Z 31/10                    & 44         \nl
766& 10-20& 2366& 2399& 2415&     &     &     & W 19$=$P 20/2/5$=$C S7/2$=$Z 36/2           & 43         \nl
769&   10 & 2377& 2400& 2402& 2410& 2420& 2428& W 13$+$A2402$+$A2428$=$Z 37/2$+$A2428$=$E 113$+$A2428    &            \nl
&      &     &     &     &     &     &     & $=$ET 193$-$A2376$-$A2448        &            \nl
801& 10-20& 2459& 2462& 2492&     &     &     & W 44$+$A2492$=$P 21/2$=$C S8/2$=$Z 38/2$=$E 118      & 45             \nl
841& 10-20& 2546& 2548& 2554&     &     &     &                                       & 12,46      \nl
852&   10 & 2572& 2589& 2592& 2593& 2657&     & W 11$-$A2506$+$A2592$=$P 22/2/5$=$C S9/2$+$A2572$=$Z 41/2$+$A2592               &            \nl
&      &     &     &     &     &     &     & $=$ E 123$+$A2592$=$ET 211$+$A2592        &            \nl
864&   10 & 2622& 2625& 2626&     &     &     & P 23/2$=$Z 43/2                                & 47         \nl
\enddata
\tablenotetext{}{
  NOTES.--(1)~In E~$6(17)$ with $z_{e}$. (2)~In ET~$10(17)$ with $z_{e}$.
  SC~17/20(2)$-$~A2734$+$~SC~52/20/40(4)$-$~A80$+$~SC~52/40/3$=$~P~13/2(9)
  $-$~A27$-$~A2716.(3)~A85 $+$~A151 $=$~T~1 $=$~B~1/2 $=$~SC~67/40/100.
  (4)~A85, A117 and A151 in W~8(7) and in ET~10(17)
  with $z_{e}$. (5)~A119 and A168 are in T2 $=$~B2(4). (6)~In W~6(9). (7)~SC~79/10/20
  $+$~SC~96/10 $=$~P~16/2 $=$~C~S4/2 $=$~Z~3/2$-$A76
  $=$~E~15 $-$~A76 $-$~A261 $=$~ET~24 $-$~A76. (8)~A154 $+$~A225
  $=$~T~3 $=$~B~3/20. (9)~In W~9(7). (10) In E~17(8) and in ET~30(8).
  (11)~SC~89/10/20 $+$~SC~130/10 $=$~P~17/2 $+$~A150 $+$~A257 $=$~Z~6/2 $=$~E~17 $=$~ET~30.
  (12)~R~$>$~300~Mpc. (13)~A225 and A257 in W~9(7). (14)~In~Z~7/2(11) with $z_{e}$
  and in E~18(11) with $z_{e}$. A266 and A277 in ET~34(6) with $z_{e}$. (15)~A3004 and
  A3009 in E~27(32) with $z_{e}$ and in ET~48(26) with $z_{e}$. (16)~In E~27(32) with $z_{e}$.
  (17)~A3093, A3100 and A3108 in C~S11/2 (15) and in C~S11/2(7). (18)~In
  ET~48(26) with $z_{e}$. (19)~In C~S11/2(7). (20)~A3202 in ET~48(26) with $z_{e}$.
  (21)~A3225 and A3266 in E~27(32) and in ET~48(26). (22)~In ET 91(9).
  (23)~In W4(10). All except A1257 in E~57(8).
  A1185 $+$~A1228 $=$~T~8 $=$~B7/20$\div$400. (24)~In ET~107(8) with $z_{e}$.
  (25)~In ET~111(16). A1307 and A1390 in E~70(8). (26)~SC~478/10 $+$~SC~486/10
  $+$~A1270 $=$~W~2(11). (27)~SC~483 and SC~543 in ET~114(16). (28)~A1291
  $+$~A1377 $+$~A1383 $+$~A1436$=$~T~9. (29)~A1365 in ET~103(2), A1423 $+$~A1480
  $=$~ET 110. (30)~In E~70(8) and in ET~111(16). (31)~A1559 and A1674 in ET~114(16).
  (32)~In Z~28/2(6). A1825 and A1827 in W~3 (10). SC 588/20$\div$100(2)
  $+$~SC~595/100$\div$400(2)$-$~A1828 $=$~P~8/2$=$C~N8/2.
  (33)~In E~83(12) and in ET~138(12). (34) A1775 and A1831 in T~12(5) and in
  B~12A(6). SC~598$-$A1898 in W~3(10).
  (35)~A1913 and A1991 in T~14(11) and in B~13/2(4). SC~624/10$+$~SC~661/10/20
  in W~1(13). (36)~In
  ET~150(10) with $z_{e}$. (37)~A2001 and A2017 in ET~150, which is at
  R$\approx$~300~Mpc, while SC~645 is at R~$=$~557~Mpc.
  (38)~A2028 and A2029 in B~14/20~(3). (39)~A2124 is not a member of our superclusters;
  A2019 $+$~A2056 $=$~SC~650/40. (40) In Z~29/2(10) and in E~90(10). SC~624/10
  $+$~SC~661/10/20 in W~1(13).
  (41)~A2107,
  A2147, A2151 and A2152 in T~14(11). (42)~In W~5(10), E~92(10) and in
  ET~160(12), which includes SC~684/20/40 $=$~A2162 $+$~A2197 $+$~A2199 having
  R~$<$~100~Mpc and SC~657/20$\div$400 $=$~A2052 $+$~A2063. SC~657/20$\div$400(2)
  $+$~SC~677/20(5) $+$~SC~684/20/40(3)$=$~P~9/2$=$~C~N9/2.
  (43)~SC~755/10/20
  $+$~SC~766/10/20$=$~E~108 $=$~ET~188 $-$~A2405e. (44)~A3806, A3822 and A3825 in
  E~109(8) and in ET~192(8) having $z_{e}$. (45)~In ET~205(19). (46)~In
  ET~209(7) with $z_{e}$. (47)~In W~10(6), E~125(6) and in ET~213(6).}
\end{deluxetable}

\clearpage

\begin{deluxetable}{rcrrrrrrrrrrrr}
\label{tbl-2}
\tabcolsep=0.1pt
\tableheadfrac{0.1}
\tablecolumns{14}
\tablenum{2}
\footnotesize
\tablecaption{SUPERCLUSTER CHARACTERISTICS}
%\tablewidth{0pt}
\tablehead{\colhead{SC} & \colhead{$f$}
& \colhead{$\nu$} & \multicolumn{1}{c}{R.A.(1950)Dec}
& \multicolumn{1}{c}{$b$} & \multicolumn{1}{c}{$R$}
& \multicolumn{1}{c}{$\Delta {\rm R.A.}$} &  \multicolumn{1}{c}{$\Delta {\rm Dec}$}
& \multicolumn{1}{c}{$\Delta R$} &  \multicolumn{1}{c}{$r_{i}$}
& \multicolumn{1}{c}{$r_{h}$} &  \multicolumn{1}{c}{$r_{s}$}
& \multicolumn{1}{c}{$r_{v}$} &  \multicolumn{1}{c}{$\cal M$} \\
\colhead{} & \colhead{}
& \colhead{} & \colhead{}
& \colhead{} & \multicolumn{1}{c}{Mpc}
& \multicolumn{1}{c}{Mpc} & \multicolumn{1}{c}{Mpc}
& \multicolumn{1}{c}{Mpc} & \multicolumn{1}{c}{Mpc}
& \multicolumn{1}{c}{Mpc} & \multicolumn{1}{c}{Mpc}
& \multicolumn{1}{c}{Mpc} & \multicolumn{1}{c}{$10^{14}{\cal M}_{\sun}$}}
\startdata
43& 10  & 4 & 0036.7$+$1849& $-$43\fdg6& 306& 22.5& 26.8& 36.1& 19.0& 24.9& 29.8& 32.5& 23.9 \nl
52& 20  & 4 & 0037.4$-$2345& $-$85.4& 195&  3.5& 11.2& 10.2&  6.5& 10.1& 10.4& 13.8& 12.2 \nl
64& 10  & 3 & 0053.3$+$0025& $-$63.0& 202& 12.7& 13.8& 19.8& 11.3& 16.7& 18.6& 24.8&  7.8 \nl
67& 10  & 4 & 0054.1$-$1231& $-$74.9& 163& 19.0& 17.3&  8.9& 10.5& 14.0& 16.1& 19.2& 13.3 \nl
75&10,20& 3 & 0101.3$-$4936& $-$67.6& 199& 11.1&  7.1&  8.0&  6.5&  9.9& 10.9& 16.3&  9.8 \nl
79&10,20& 3 & 0104.0$+$0007& $-$62.2& 134& 11.0&  8.1&  3.1&  5.8&  9.3&  9.8& 15.8& 11.4 \nl
89& 10  & 4 & 0109.6$+$1543& $-$46.6& 196&  6.3& 15.4& 32.4& 13.5& 17.3& 20.5& 22.6& 12.5 \nl
& 20  & 3 & 0108.0$+$1537& $-$46.7& 190&  2.0& 14.9& 17.3&  9.6& 13.7& 15.7& 21.3& 10.0 \nl
96& 10  & 3 & 0119.0$+$1413& $-$47.7& 136&  8.0& 25.1& 18.7& 13.3& 20.1& 22.0& 33.3&  7.4 \nl
117& 10  & 3 & 0136.1$-$0830& $-$68.0& 349&  8.9& 23.2& 12.9& 11.7& 17.8& 19.4& 26.4& 10.9 \nl
121& 10  & 3 & 0140.3$-$1101& $-$69.6& 396& 19.2& 13.4&  5.1& 10.9&  6.4& 16.3&  8.9&  8.3 \nl
130& 10  & 4 & 0152.1$+$1740& $-$42.3& 207& 26.1& 20.8& 15.2& 14.6& 16.4& 22.4& 24.1& 10.7 \nl
131&10,20& 3 & 0152.7$-$0602& $-$63.8& 284&  5.5& 16.0& 21.6& 12.2& 20.5& 20.9& 31.0&  8.1 \nl
159&10,20& 3 & 0216.1$-$4807& $-$63.3& 197&  5.3&  4.9&  7.7&  4.4&  7.4&  7.5& 11.6&  9.7 \nl
203&10,20& 3 & 0308.6$-$2606& $-$58.9& 203&  3.2& 12.4&  9.9&  7.1& 11.9& 12.2& 18.4&  9.7 \nl
206& 10  & 4 & 0312.5$-$4651& $-$55.8& 189&  3.2& 13.0& 14.5&  8.3&  4.3& 12.4&  6.3& 14.6 \nl
& 20  & 3 & 0313.6$-$4637& $-$55.7& 192&  1.6& 13.3&  0.6&  6.2&  2.5&  9.0&  3.0&  9.1 \nl
225& 10  & 4 & 0335.3$-$5421& $-$49.6& 181&  8.6& 14.1&  9.2&  7.3&  8.2& 11.0& 10.5& 18.2 \nl
& 20  & 3 & 0332.2$-$5323& $-$50.4& 179&  7.3&  3.3&  4.0&  3.9&  5.5&  6.4&  9.6& 16.3 \nl
241& 10  & 3 & 0350.6$-$5349& $-$47.6& 116&  6.9&  5.5& 26.9& 11.7& 17.4& 19.2& 25.3&  9.6 \nl
258&10,20& 3 & 0416.8$-$6201& $-$41.5& 174&  7.9& 15.2& 13.9&  9.1& 15.6& 15.7& 24.2& 11.6 \nl
270& 10  & 3 & 0436.9$-$1009& $-$33.8& 112& 25.3& 11.0& 20.4& 14.6& 24.8& 25.1& 40.6& 11.8 \nl
315&10,20& 3 & 0838.6$+$2758&  35.2& 259&  5.4&  9.5& 34.0& 14.7& 21.2& 23.9& 32.3&  8.4 \nl
332& 20  & 4 & 0919.1$+$7355&  36.3& 414&  4.8& 15.2&  7.3&  7.2&  8.4& 10.9& 12.6& 17.8 \nl
343& 10  & 4 & 0933.5$+$7528&  36.3& 379&  7.9& 21.9& 12.7& 10.2& 12.6& 15.3& 15.8& 10.0 \nl
& 20  & 3 & 0933.5$+$7446&  36.7& 376&  8.2&  5.8&  8.2&  5.6&  9.1&  9.5& 13.6&  8.2 \nl
377& 20  & 3 & 1017.8$+$5049&  53.1& 415& 12.0& 16.1&  5.7&  9.5& 15.8& 16.2& 24.1& 10.0 \nl
378& 10  & 6 & 1018.7$-$0707&  39.8& 172& 24.8& 17.3& 27.1& 13.5& 14.2& 18.9& 17.6& 16.7 \nl
& 20  & 5 & 1015.0$-$0654&  39.4& 169& 11.0& 17.0& 18.8&  9.8& 11.4& 14.2& 14.9& 14.0 \nl
379&10,20& 3 & 1018.9$+$5019&  53.5& 486&  8.5&  2.4& 11.6&  6.0& 10.2& 10.3& 15.6& 10.4 \nl
419& 10  & 3 & 1106.6$+$3849&  65.7& 237&  5.8& 23.4& 17.4& 13.0& 11.6& 19.9& 15.5& 11.1 \nl
420& 10  & 3 & 1108.6$+$0410&  56.5& 223& 19.5& 25.5& 13.0& 14.6& 23.9& 24.8& 36.7&  8.4 \nl
431&10,20& 5 & 1116.5$+$2940&  69.0& 101&  7.1& 24.5& 15.5& 10.8& 13.7& 16.0& 17.1& 12.8 \nl
435& 10  & 5 & 1120.6$+$7525&  40.7& 373& 21.7& 16.6& 26.3& 12.9& 18.3& 19.8& 26.7& 18.3 \nl
& 20  & 4 & 1114.3$+$7524&  40.5& 377& 15.7& 16.8& 14.6& 10.6& 16.0& 16.8& 24.0& 12.8 \nl
448&10,20& 3 & 1127.3$-$0358&  52.9& 159& 14.7&  1.4& 21.2& 10.8& 18.3& 18.5& 27.5&  8.0 \nl
462& 10  & 3 & 1134.1$+$6119&  53.8& 536& 22.0& 14.0& 37.4& 19.6& 30.5& 32.7& 45.5&  7.5 \nl
470&10,20& 4 & 1139.4$+$1056&  66.8& 333&  7.0&  8.4& 23.9& 10.3& 10.4& 15.2& 15.1& 15.0 \nl
473&10,20& 4 & 1139.9$+$1224&  67.9& 254& 18.2& 19.5&  2.5& 10.2& 10.3& 15.4& 16.8& 12.2 \nl
478& 10  & 3 & 1142.6$+$5301&  61.3& 246& 30.5& 15.8& 19.6& 16.9& 28.4& 29.0& 43.6&  8.1 \nl
483&10,20& 5 & 1147.0$+$6749&  48.5& 358& 15.7& 10.5& 16.4&  8.8& 10.2& 13.1& 12.6& 19.4 \nl
486& 10  & 7 & 1149.3$+$5553&  59.4& 177& 19.2& 25.7& 40.0& 17.2& 20.0& 24.4& 22.9& 23.0 \nl
& 20  & 3 & 1135.7$+$5551&  58.6& 162&  6.0&  3.1& 16.3&  7.4& 10.9& 12.2& 16.2& 10.6 \nl
493&10,20& 3 & 1154.9$+$3204&  77.3& 230& 22.5& 11.2&  9.3& 11.4& 18.7& 19.4& 29.3& 12.1 \nl
524& 10  & 4 & 1221.5$+$1431&  75.5& 247& 29.7& 20.9& 20.2& 15.4& 22.6& 24.4& 30.3& 14.8 \nl
537& 10  & 4 & 1240.7$+$6919&  48.0& 333& 15.8& 29.8& 14.2& 14.8& 21.7& 23.3& 36.1& 18.0 \nl
543& 20  & 3 & 1244.4$+$6321&  54.0& 317& 13.0& 12.3& 19.8& 11.0& 16.2& 18.0& 27.9& 10.3 \nl
558&10,20& 3 & 1301.5$+$3110&  85.3& 521&  9.3& 24.9& 16.4& 14.0& 22.9& 23.8& 34.2& 18.6 \nl
560& 10  & 3 & 1302.9$+$3316&  83.3& 579&  2.3& 36.9& 16.2& 17.1& 25.4& 28.4& 46.5& 12.9 \nl
587&10,20& 3 & 1344.2$+$0340&  62.8& 240& 11.7& 12.2&  4.4&  7.5&  8.8& 11.8& 13.8& 12.9 \nl
595& 20  & 3 & 1355.8$+$2030&  73.2& 195&  0.3& 11.3& 14.8&  7.9& 11.6& 12.9& 18.4& 10.6 \nl
598& 10  & 5 & 1358.3$+$2723&  74.6& 230& 34.8& 12.0& 15.5& 14.7& 17.3& 21.3& 23.2& 17.6 \nl
623& 10  & 4 & 1431.4$+$5613&  55.8& 422&  8.4& 25.5& 24.2& 14.7& 10.5& 22.2& 15.1& 24.1 \nl
624& 10  & 3 & 1431.9$+$1753&  64.4& 167& 23.2&  5.7& 18.2& 13.2& 11.7& 20.3& 21.0&  8.8 \nl
635&10,20& 4 & 1443.5$+$5525&  55.1& 318& 20.6& 14.1& 15.1& 12.2&  4.4& 17.8&  6.2& 15.1 \nl
640&10,20& 6 & 1450.5$+$2152&  61.9& 360& 16.3& 27.0& 11.7& 10.9& 12.8& 15.5& 15.3& 18.4 \nl
641&10,20& 3 & 1452.8$+$2808&  62.7& 386&  9.5&  1.9& 12.1&  6.5& 10.9& 11.1& 16.4& 20.3 \nl
645&10,20& 3 & 1457.6$+$2313&  60.7& 557& 12.3&  3.9& 19.4&  9.6& 13.6& 15.8& 19.4& 13.4 \nl
648& 20  & 3 & 1508.2$+$0643&  51.1& 240&  2.1&  7.4& 15.9&  7.5& 11.6& 12.6& 17.3& 10.3 \nl
661& 10  & 6 & 1524.8$+$2943&  55.8& 219& 10.1& 13.2& 37.8& 15.1& 17.4& 21.6& 20.7& 25.2 \nl
   & 20  & 4 & 1522.9$+$2930&  56.2& 228&  9.9& 12.7& 18.0&  9.8& 12.9& 15.1& 18.2& 18.5 \nl
665&10,20& 3 & 1533.5$+$3731&  54.2& 475&  3.5&  5.0&  2.9&  2.8&  4.2&  4.6&  6.0& 21.4 \nl
677& 20  & 5 & 1557.0$+$1937&  46.4& 115& 12.1& 19.3& 20.8& 11.1&  9.3& 15.8&  9.7& 17.5 \nl
679& 10  & 3 & 1611.3$+$2722&  45.4& 285& 25.7& 26.3& 24.4& 18.8& 32.1& 32.4& 50.5& 12.1 \nl
681&10,20& 3 & 1614.8$+$5116&  44.8& 178&  3.8& 15.6& 19.7& 10.7& 14.5& 17.5& 25.6&  8.5 \nl
683&10,20& 6 & 1619.7$+$4216&  45.0& 420& 35.1& 15.4& 17.5& 13.3& 13.4& 18.7& 15.7& 19.3 \nl
699& 10  & 3 & 1707.4$+$3537&  35.3& 258& 11.4& 23.0& 25.6& 15.3& 22.6& 25.4& 34.1&  8.8 \nl
702&10,20& 4 & 1720.7$+$7755&  31.3& 183&  9.5&  5.4& 29.3& 11.8& 14.4& 17.8& 18.6& 11.2 \nl
755& 10  & 5 & 2144.5$-$1706& $-$46.3& 183& 15.1& 18.8& 22.2& 12.9& 12.3& 19.0& 14.6& 16.0 \nl
& 20  & 3 & 2141.2$-$1458& $-$44.8& 189& 10.3&  4.6& 12.0&  7.2&  6.5& 11.2&  9.4& 10.1 \nl
763&10,20& 4 & 2151.2$-$5808& $-$46.5& 229&  7.1& 17.0&  5.0&  7.0&  9.6& 10.9& 12.2& 22.0 \nl
766&10,20& 3 & 2152.6$-$0658& $-$43.5& 174& 17.1&  6.7& 14.8&  9.9& 14.3& 16.3& 21.5&  8.3 \nl
769& 10  & 6 & 2159.2$-$1040& $-$46.8& 254& 33.1& 12.5& 23.4& 14.1& 16.1& 20.7& 22.8& 22.5 \nl
801&10,20& 3 & 2239.4$-$1741& $-$58.6& 214& 12.4& 13.4&  5.3&  8.0& 10.8& 12.9& 17.0&  7.1 \nl
841&10,20& 3 & 2308.8$-$2148& $-$66.6& 341&  2.4& 13.3&  5.7&  6.0&  8.9&  9.8& 13.4& 18.8 \nl
852& 10  & 5 & 2324.7$+$1513& $-$42.7& 127& 14.1& 21.4& 19.3& 11.4& 12.2& 16.4& 16.4& 12.6 \nl
864& 10  & 3 & 2333.4$+$2246& $-$36.6& 180&  1.2& 21.8& 14.8& 11.7& 17.3& 19.5& 25.4&  8.0 \nl
\enddata
\end{deluxetable}

\clearpage

\begin{deluxetable}{rrlccc}
\label{tbl-3}
\tableheadfrac{0.1}
\tablecolumns{6}
\tablenum{3}
\footnotesize
\tablecaption{APM SUPERCLUSTERS OF GALAXIES}
\tablewidth{0pt}
\tablehead{
\colhead{} & \colhead{}
& \colhead{} & \colhead{}
& \colhead{} & \multicolumn{1}{c}{$R$}\\ [-5pt]
\multicolumn{1}{c}{APMS} & \multicolumn{1}{c}{$\nu$}
& \multicolumn{1}{c}{APMC (A, S)} & \multicolumn{1}{c}{R.A.(1950)Dec}
& \multicolumn{1}{c}{$b$} & \colhead{}\\ [-5pt]
\colhead{} & \colhead{}
& \colhead{} & \colhead{}
& \colhead{} & \multicolumn{1}{c}{Mpc}}
\startdata
1 &4&  1(A2717), 5, 12, 933                        & 2359.6$-$3516& $-$77.2& 149 \nl
14 &3&  112(A2829), 119(A118), 130(A122)            & 0051.4$-$2606& $-$88.4& 348 \nl
18 &3&  160(S144), 162, 173(A2911)                  & 0118.4$-$3705& $-$78.6& 235 \nl
21 &4&  182(S160), 193, 194(S167), 209(S186)        & 0133.9$-$3329& $-$79.0& 206  \nl
23 &4&  204(A2933), 211, 213, 214                   & 0142.7$-$5541& $-$61.4& 285  \nl
28 &3&  253(S239), 255, 257(A3004)                  & 0215.7$-$4843& $-$63.9& 194  \nl
29 &3&  268, 269, 270(A3027)                        & 0227.8$-$3459& $-$68.0& 235  \nl
38 &9&  330(A3078), 380(S339), 395(A3125), 396, 399,& 0327.2$-$5313& $-$51.3& 181  \nl
& &  403(A3128), 421(S366), 434(A3158), 445      &            &      &       \nl
40 &3&  345(A3094), 349(A3095), 359(S333)           & 0311.0$-$2837& $-$58.6& 203   \nl
62 &5&  642, 650, 653(A3757), 657(S933), 659        & 2114.9$-$4543& $-$44.4& 299   \nl
64 &4&  688, 700(S96), 709(A3809), 711(S974)        & 2140.3$-$4342& $-$49.1& 193   \nl
73 &3&  774(S1022), 813(A3907), 822(A3921)          & 2236.8$-$6441& $-$48.2& 294   \nl
75 &4&  811, 812, 814(A3908), 815(A3910)            & 2242.0$-$4513& $-$59.3& 279   \nl
77 &3&  830(A3925), 844(S1080), 854(A3972)          & 2256.9$-$4554& $-$62.0& 263  \nl
82 &3&  902, 904(A4010), 911                        & 2329.4$-$3634& $-$71.3& 293  \nl
83 &3&  905(A4012), 917, 920(A2660)                 & 2335.7$-$3057& $-$73.8& 157  \nl
\enddata
\end{deluxetable}

\clearpage

\begin{deluxetable}{cccrcc}
\label{tbl-4}
\tablecolumns{6}
\tablenum{4}
\tablecaption{Data for samples of superclusters}
\tablehead{
\colhead{Sample} & \colhead{Catalog} & \colhead{Cap(s)}  &
\colhead{$f$}
    & \colhead{$n$}   & \colhead{Distance}}
\startdata
1 & KK      & N+S    & 10     & 61    & $z$\nl
2 & KK      & N      & 10     & 35    & $z$\nl
3 & KK      & S      & 10     & 26    & $z$\nl
4 & KK      & N+S    & 20     & 49+30 & $z+(e)$\nl
5 & ZZSR    & N+S    & 2      & 23+35 & $z+e$\nl
6 & EETDA   & N+S    & 2      & 21+31 & $z+e$\nl
7 & ETJEA   & N+S    & 2      & 29+42 & $z+e$\nl
8 & APMS    & S      & $\approx 10$ & 83+42 & $z+e$\nl
9 & APMS    & S      & $\approx 10$ & 16    & $z$\nl
\enddata
\end{deluxetable}

\clearpage

\epsfxsize=8cm 

\mbox{\epsfxsize=8cm \epsfbox{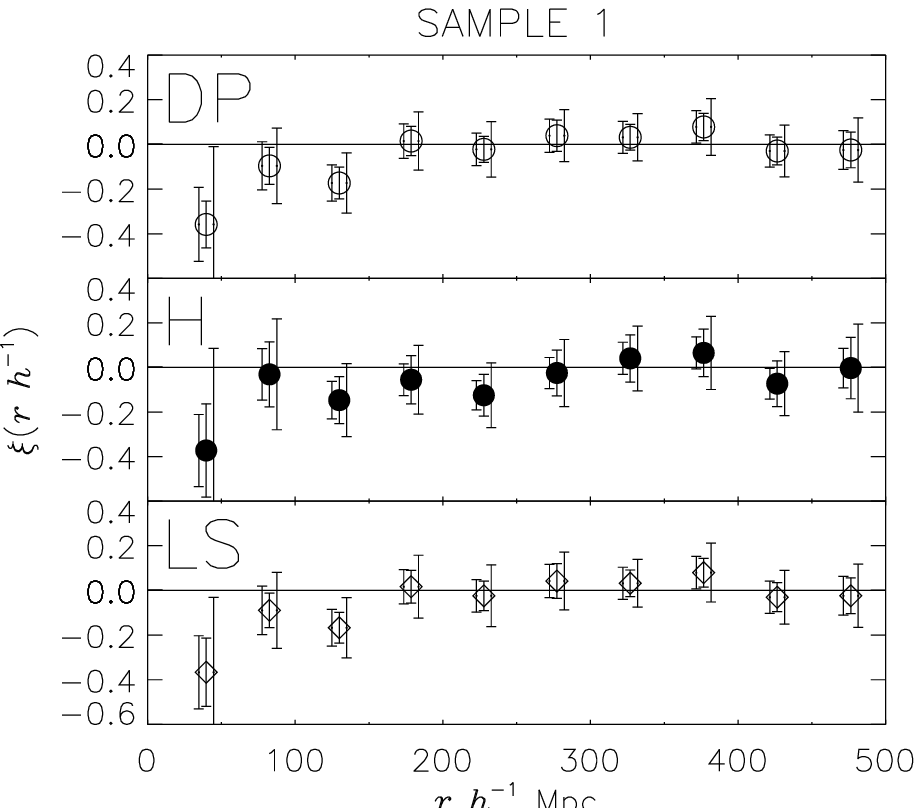} \epsfxsize=8cm \epsfbox{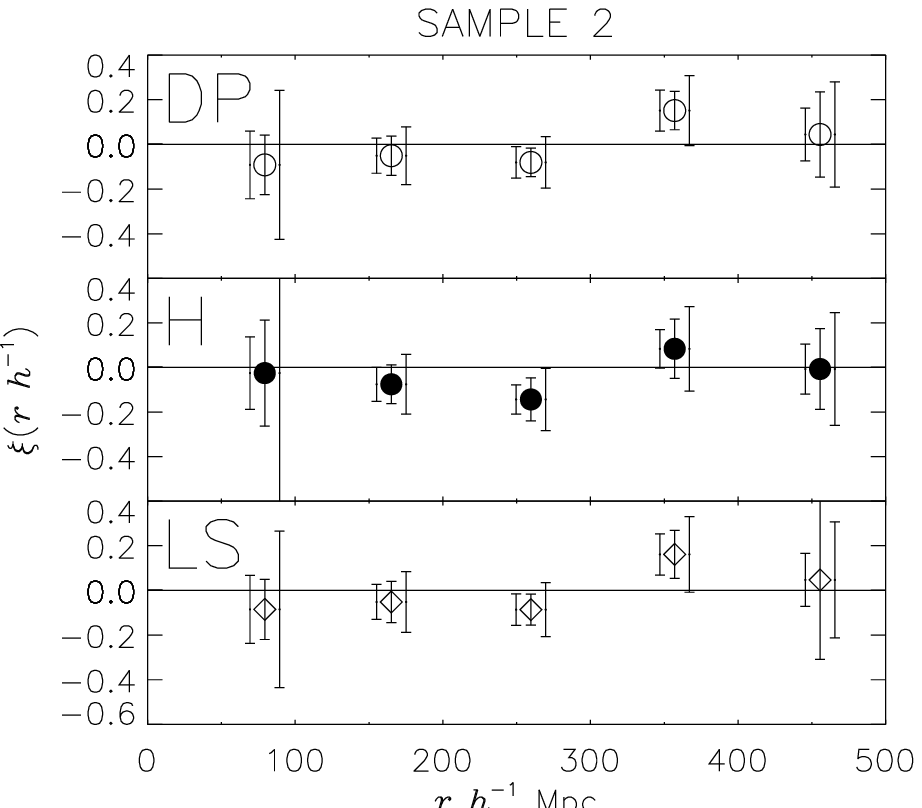}}

\mbox{\epsfxsize=8cm \epsfbox{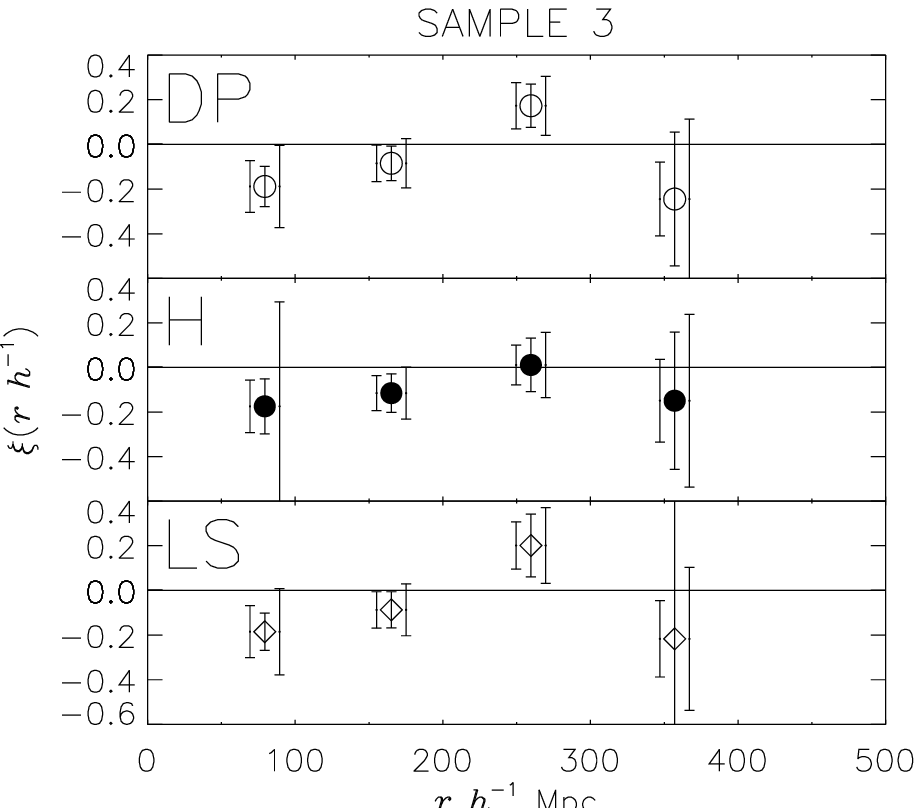} \epsfxsize=8cm \epsfbox{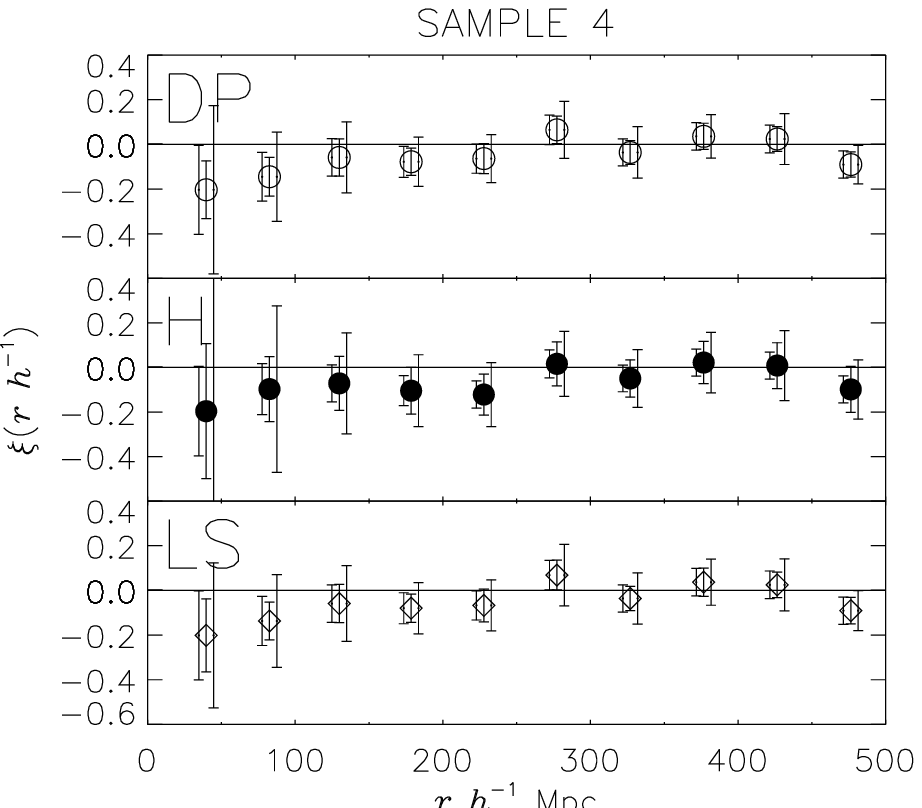}}

\clearpage

\epsfxsize=8cm \epsfbox{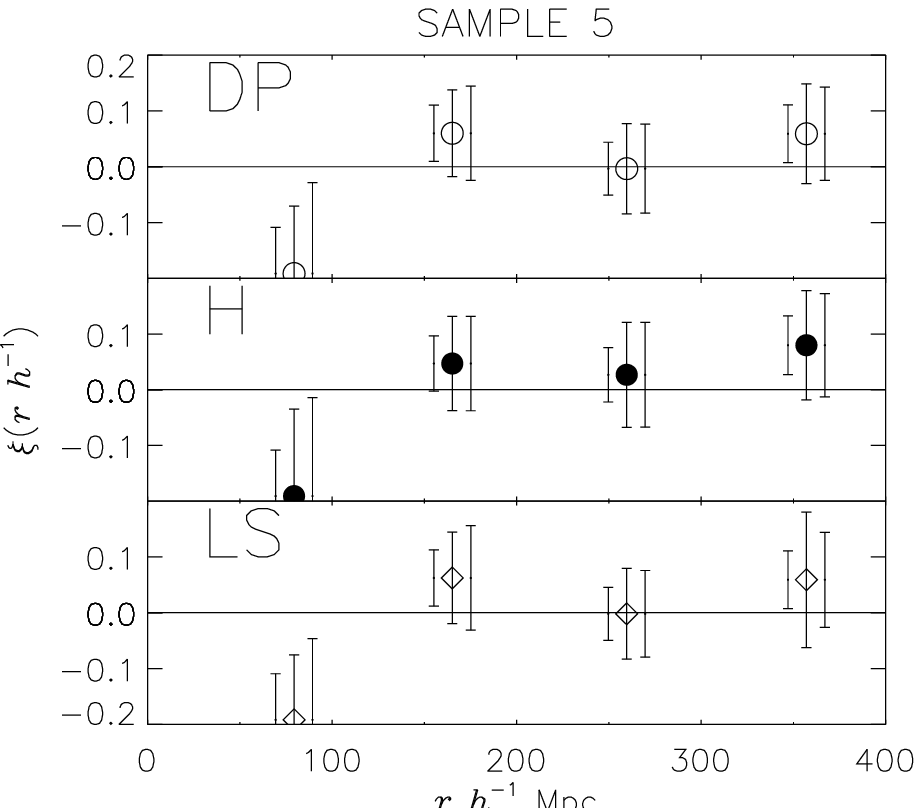} 

\epsfxsize=8cm \epsfbox{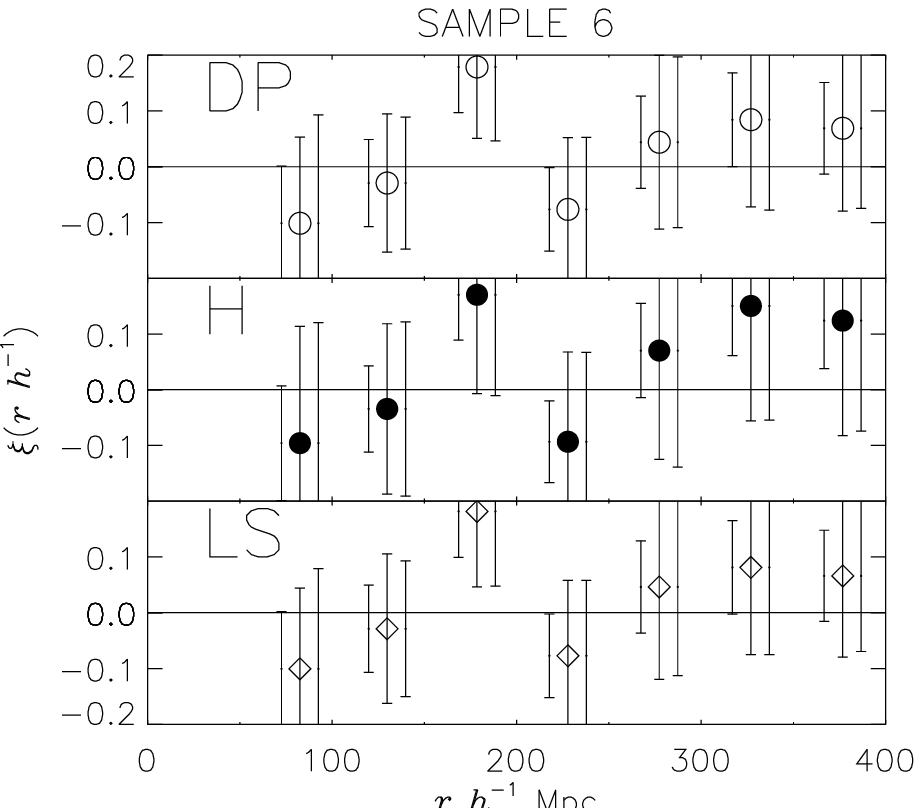}

\epsfxsize=8cm \epsfbox{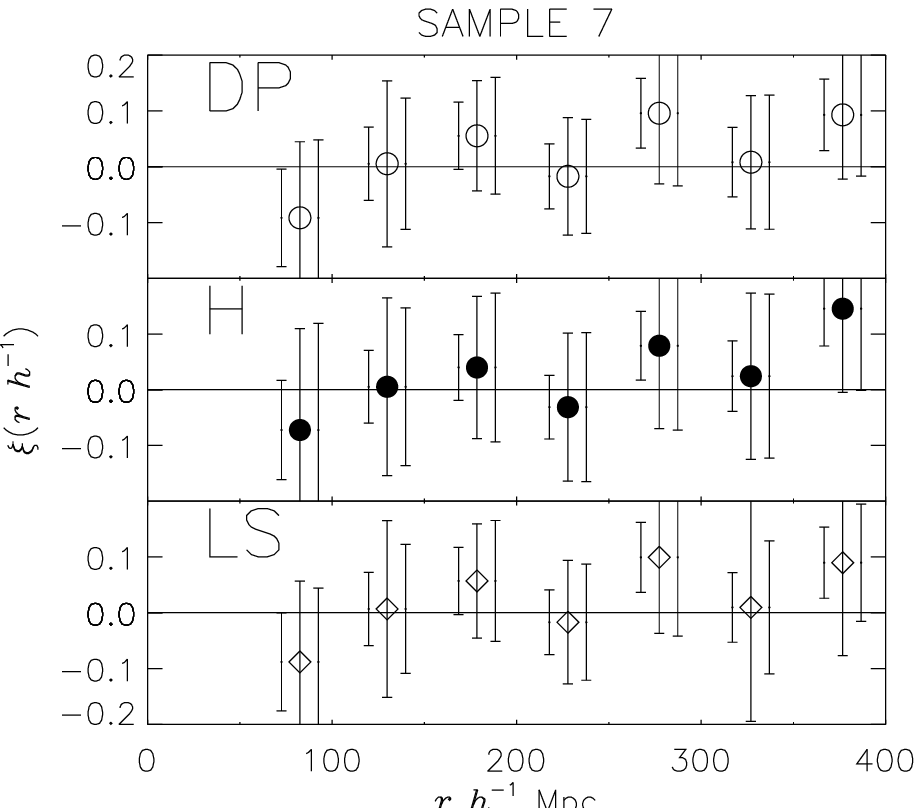}

\clearpage

\epsfxsize=8cm \epsfbox{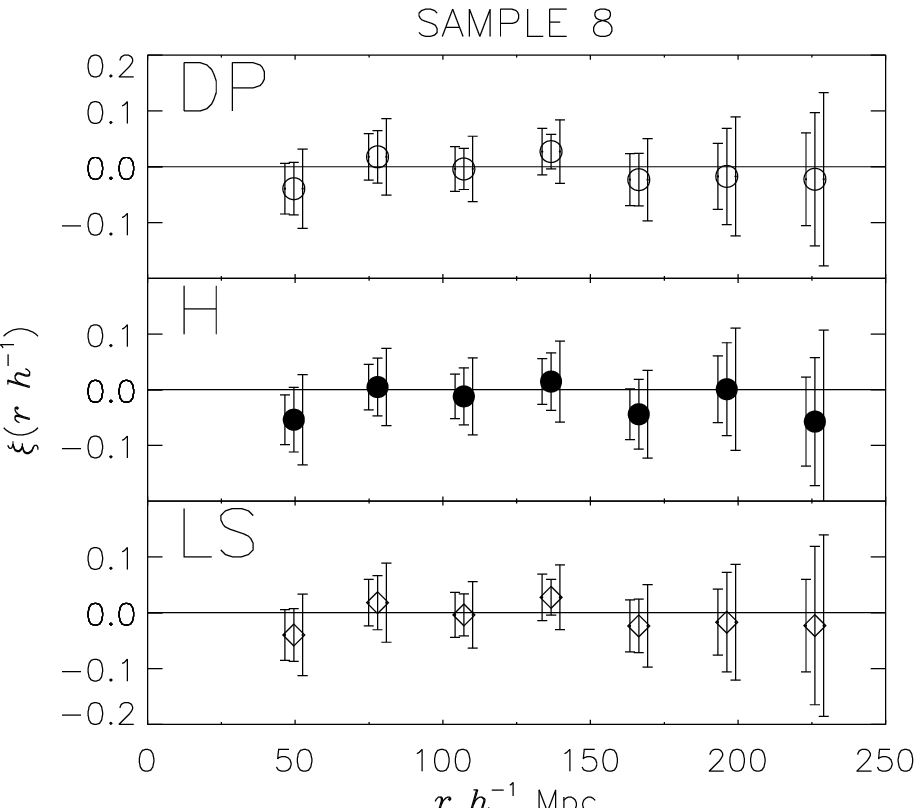} 
\epsfxsize=8cm \epsfbox{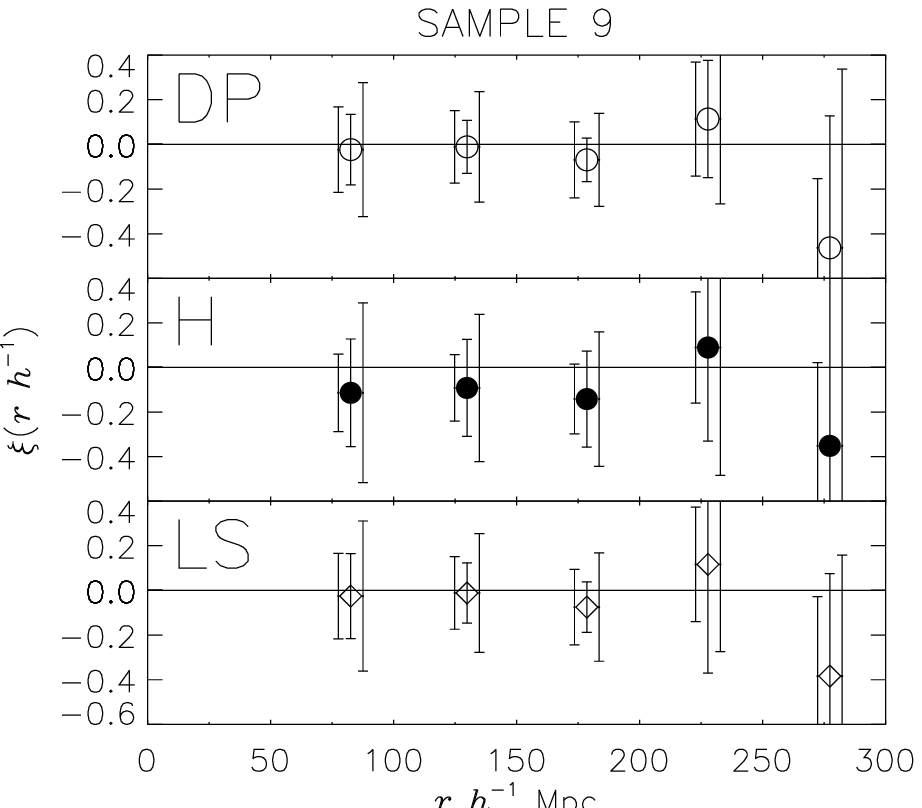}


\clearpage

\begin{thebibliography}{}
\bibitem[]{ab58} Abell, G. O. 1958, \apjs, 3, 211
\bibitem[]{aco89} Abell, G. O., Corwin H. G., \& Olowin, R. P. 1989,
  \apjs, 70, 1
\bibitem[]{b88} Bahcall, N. A. 1988, \araa, 26, 631
\bibitem[]{bb86} Bahcall, N. A., \& Burgett, W. S. 1986, \apj, 300, L35
\bibitem[]{bc93} Bahcall, N. A., \& Cen, R. 1993, \apj, 407, L49
\bibitem[]{bs86} Bahcall, N. A., \& Soneira, R. M. 1984, \apj, 277, 28
\bibitem[]{bw92} Bahcall, N. A., \& West, M. J. 1992, \apj, 392, 419
\bibitem[]{bb85} Batuski, D. J. \& Burns, J. O. 1985, AJ, 90, 1413
\bibitem[]{ber94} Bernstein, G. M. 1994, \apj, 424, 569
\bibitem[]{cm92} Cappi, A. \& Maurogordato, S. 1992, \aap, 259, 423 (C)
\bibitem[]{dmse97} Dalton, G. B., Maddox, S. J., Sutherland, W. J., \&
  Efstathiou, G. 1997, \mnras, 289, 263
\bibitem[]{dp83} Davis, M., \& Peebles, P. J. E. 1983, \apj, 267, 465
\bibitem[]{et86} Efron, B., \& Tibshirani, R. 1986, Stat.  Sci., 1, 54
\bibitem[]{e92} Efstathiou, G., Dalton, G. B., Sutherland, W. J., \&
  Maddox, S. J. 1992, \mnras, 257, 125
\bibitem[]{eetda} Einasto, M., Einasto J., Tago, E., Dalton, G. B., \&
  Andernach, H. 1994, \mnras, 269, 301 (EETDA,E)
\bibitem[]{etjea} Einasto, M., Tago, E., Jaaniste, J., Einasto, J., \&
  Andernach, H. 1997, \aaps, 123, 119 (ETJEA,ET)
\bibitem[]{f94} Fisher, K. B., Davis, M., Strauss, M. A., Yahil, A.,
  \& Huchra, J. 1994, \mnras, 266, 50
\bibitem[]{gcf92} Gourgoulhon, E., Chamaraux, P., \& Fouqu\'{e},
  P. 1992, \aap, 255, 69
\bibitem[]{h93} Hamilton, A. J.  S. 1993, \apj, 417, 19
\bibitem[]{h96} Hermit, S., Santiago, B. X., Lahav, O., Strauss,
  M. A., Davis, M., Dressler, A., \& Huchra, J. P. 1996,
  \mbox{\mnras}, 283, 709
\bibitem[]{hg82} Huchra, J. P., \& Geller, M. J. 1982, \apj, 257, 423
\bibitem[]{j75} Jackson, J. C. 1975, \mnras, 173, 41p
\bibitem[]{kk85} Kalinkov, M., \& Kuneva, I.  1985, Astron. Tsirk.
  (Moscow), No. 1409, 1
\bibitem[]{kk86} \rule[0.5ex]{4em}{0.4pt}. 1986, \mnras, 218, 49p
\bibitem[]{kk95}  \rule[0.5ex]{4em}{0.4pt}. 1995, \aap, 113, 451 (KK)
\bibitem[]{kkv94} Kalinkov, M., Kuneva, I., \& Valtchanov, I. 1994, in
  Astronomical Data Analysis Software and Systems III, ed. D. R.
  Crabtree, R. J. Hanish and J. Barnes (ASP Conf. Ser. Vol. 61), p. 263
  (San Francisco)
\bibitem[]{kvk96} Kalinkov, M., Valtchanov, I., \& Kuneva, I. 1996, in
  Proc. 2nd Hellenic Astron. Conf., ed. M. E. Contadakis,
  J. D. Hadjidemetriou, L. N. Mavridis, J. H. Seiradakis,
  p. 364 (Thessaloniki)
\bibitem[]{kvk97} \rule[0.5ex]{4em}{0.4pt}. 1998, \aap, 331, 838
\bibitem[]{ls} Landy, S. D., \& Szalay, A. 1993, \apj, 412, 64
\bibitem[]{ll88} Lebedev, V. S., \& Lebedeva, I. A.  1988, Letters to
  AZh, 14, 18
\bibitem[]{lfb86} Ling, E. N., Frenk, C. S., \& Barrow, J. D. 1986,
  \mnras, 223, 21p
\bibitem[]{l95} Loveday, J., Maddox, S. J., Efstathiou, G., \&
  Peterson, B. A. 1995, \apj, 442, 457
\bibitem[]{msel90} Maddox, S. J., Sutherland, W. J., Efstathiou, G.,
  \& Loveday, J. 1990a, \mnras, 243, 692
\bibitem[]{mes90} Maddox, S. J., Efstathiou, G., \& Sutherland, W. J.
  1990b, \mnras, 246, 433
\bibitem[]{mes96} Maddox, S. J., Efstathiou, G., \& Sutherland, W. J.
  1996, \mnras, 283, 1227
\bibitem[]{mdcl89} Maia, M. A. G., da Costa, L. N., \& Latham,
  D. W. 1989, \apjs, 69, 809 
\bibitem[]{mjb92} Mo, H. J., Jing, Y. P., \& B\"{o}rner, G. 1992,
  \apj, 392, 452
\bibitem[]{nw87} Nolthenius, R., \& White, S. D. M. 1987, \mnras, 235, 505
\bibitem[]{pew92} Peacock, J. A., \& West, M. J. 1992, \mnras, 253, 307
\bibitem[]{pee80} Peebles, P. J. E. 1980, The Large-Scale Structure of the
  Universe. Princeton, Princeton Univ. Press
\bibitem[]{po86} Postman, M., Geller, M. J., \& Huchra, J. P. 1986,
  \aj, 91, 1267
\bibitem[]{phg92} Postman, M., Huchra, J. P., \& Geller, M. J.  1992,
  \apj, 384, 404 (P)
\bibitem[]{phgh85} Postman, M., Huchra, J. P., Geller, M. J., \& Henry, J. P.
  1985, AJ, 90, 1400
\bibitem[]{pssj89} Postman, M., Spergel, D. N., Sutin, B., \& Juszkiewicz, R.
  1989, \apj, 346, 588
\bibitem[]{rgh89} Ramella, M., Geller, M. J., \& Huchra, J. P. 1989,
  \apj, 344, 57
\bibitem[]{szvc91} Scaramella, R., Zamorani, G., Vettolani, G., \&
  Chincarini, G. 1991, AJ, 101, 342
\bibitem[]{scye96} Shepherd, C. W., Carlberg, R. G., Yee, H. K. C., \&
  Ellingson, E. 1996 (preprint)
\bibitem[]{s95} Sicotte, H. 1995, Ph. D. thesis, Princeton University
\bibitem[]{ss85} Szalay, A. S., \& Schramm, D. N. 1985, Nature, 314,
  718
\bibitem[]{th80} Thuan, T. X. 1980, in Physical Cosmology. Les Houches,
  Session 32, ed. R. Balian, J. Audouze, D. N. Schramm, North-Holland, p. 277
\bibitem[]{tu87} Tully, R. B. 1987, \apj, 323, 1
\bibitem[]{we89} West, M. J. 1989, \apj, 347, 610 (W)
\bibitem[]{wb91} West, M. J. \& van den Bergh, S. 1991, \apj, 373, 1
\bibitem[]{zzsv} Zucca, E., Zamorani, G., Scaramella, R., \&
  Vettolani, G.  1993, \apj, 407, 470 (ZZSV)
\end{thebibliography}
\end{document}